# Real-time Emissions of Telecom Core Networks

Hatem A. Alharbi, Mohamed Musa, Taisir E. H. El-Gorashi and Jaafar M. H. Elmirghani
*School of electronic and electrical engineering, University of Leeds, United Kingdom*

**ABSTRACT**

This paper focuses on investigating the direct and life cycle assessment (LCA) emissions in core networks, considering British telecom (BT) core network as a case study. We found that on 1st December 2016, the total BT core network emissions were around 22 $tCO_2e$ and 25 $tCO_2e$ of direct and LCA emissions, respectively.

Keywords: **IP over WDM network, energy consumption, direct emission, LCA emission and MILP.**

## 1. INTRODUCTION

Energy consumption of information and communication technology (ICT) is continuously gaining mounting attention. By 2020, ICT is projected to be the source of 14% of the global energy consumption [1] and the communication network is a major contributor to ICT energy consumption as it is expected to be responsible for 12% of the total ICT energy consumption [2]. The volume of Internet traffic and consequently the energy consumption resulting from it are increasing dramatically. In 2016 [3], the United Kingdom's IP networks carried 129 Petabytes per day with 88% of this traffic originating from content delivery datacentres and by 2021, the total IP traffic in the United Kingdom is expected to grow 3-times its value in 2016 reaching 10.6 Exabytes per month. In the present research, we focus on estimating the energy consumption of IP over WDM core networks as they provide a high network bandwidth for the Internet backbone, transporting traffic. In the literature, several papers have a focus on minimizing the energy consumption of IP over WDM networks considering multiple dimensions [4]–[15].

The emissions attributed to electricity production and the associated Greenhouse gases (GHG) have increased the global interest to tackle the problem, and the first step towards reducing these emissions lies in formulating models to measure them. According to the Intergovernmental Panel on Climate Change (IPCC), electricity production is responsible for approximately 40% of the global emissions [16] and fossil fuels contribute 68% of the global energy consumption [17]. In order to calculate the emissions of electricity production, two factors are used; direct [18] and life cycle assessment [19] (LCA) emissions. Direct emission refers to the emission happening during the electricity generation, whereas, LCA emission refers to the emission taking place throughout the life cycle of electricity production.

This paper aims to estimate the direct and LCA emissions of core networks, by considering BT core network as an example with its associated traffic demands and equipment power consumption as well as the emission factors in the UK. The rest of this paper is organized as follows: In Section 2, we discuss Great Britain (GB) electricity national grid and its direct and LCA carbon intensity factors. In Section 3, we calculate the energy consumption of BT core network and the associated direct and LCA emissions. Finally, the paper is concluded in Section 4.

## 2. GB National Grid Electricity Production

Fig. 1 shows the GB national grid electricity production by fuel type and the total demand, which are obtained from [20], [21], at 5 minute periods for selected days in 2016 (1st January in (a), 1st April in (b), 1st August in (c) and 1st December in (d)). Among all electricity generations, the Combined Cycle Gas Turbines (CCGT) is the major electricity generation source in the selected days of GB national grid, whereas; nuclear sources generate large amounts of consistent electricity over the selected days.

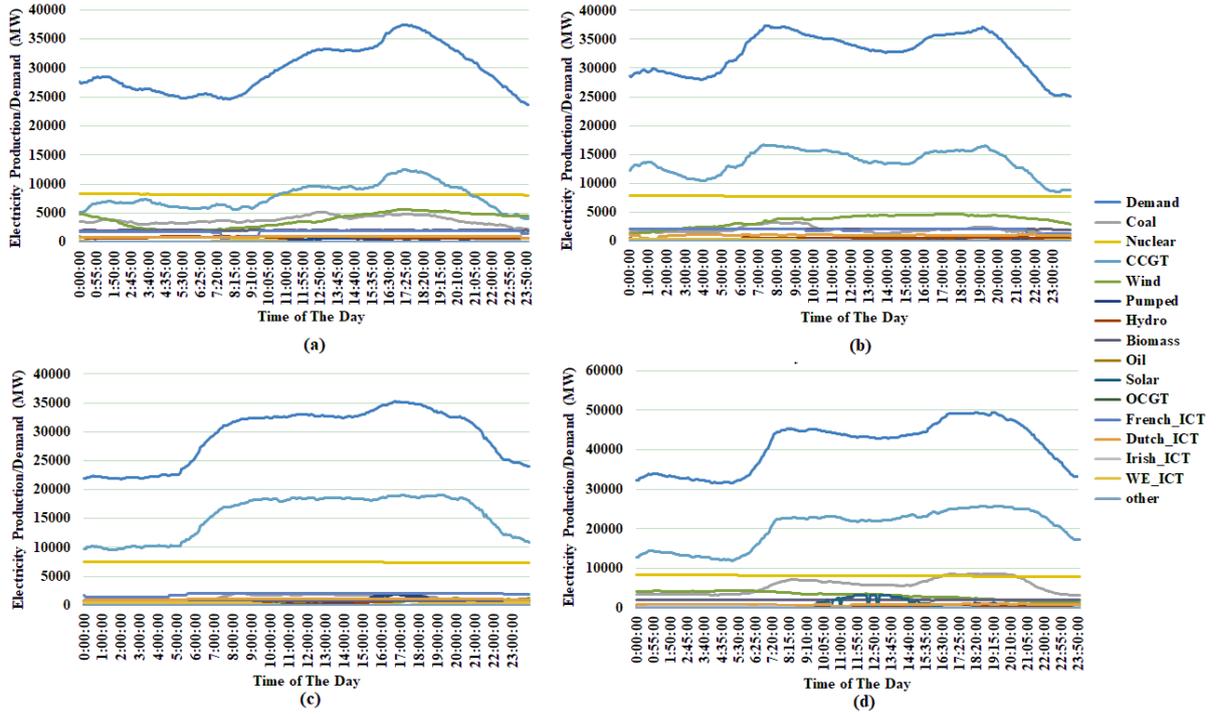

Figure 1: GB electricity production by fuel type and the total electricity demand in a) 1st January, b) 1st April, c) 1st August and d) 1st December.

Carbon emission intensity [22] is the carbon emission per unit of energy consumed. The common carbon emission intensity unit is kgCO2e/kWh. Table 1 shows the direct and LCA carbon intensity factors for each GB fuel type. For the purpose of measuring the direct and LCA emissions associated with GB national grid (Fig. 1), the GB carbon intensity at a given time is calculated using [18]:

$$C_t = \frac{P_g \, \varphi_g}{D_t} \quad (1)$$

Equation (1) calculates the carbon intensity $C_t$ at time $t$; where $P_g$ is the generation production of fuel type $g$, $\varphi_g$ is the carbon intensity (direct or LCA) of fuel source $g$ (see table 1) and $D_t$ is the total electricity demand at time $t$.

Table 1: Direct and LCA Carbon intensity factors for each GB fuel type [17], [22]

| Generation Type | Direct Carbon Intensity (kgCO2e/kWh) | LCA Carbon Intensity (kgCO2e/kWh) |
|---|---|---|
| Coal | 0.937 | 0.82 |
| Nuclear | 0 | 0.012 |
| CCGT | 0.394 | 0.49 |
| Wind | 0 | 0.012 |
| Pumped | 0 | 0 |
| Hydro | 0 | 0.024 |
| Biomass | 0.12 | 0.23 |
| Oil | 0.935 | 0.935 |
| Solar | 0 | 0.048 |
| OCGT | 0.651 | 0.651 |
| French ICT | 0.053 | 0.053 |
| Dutch ICT | 0.474 | 0.474 |
| Irish ICT | 0.458 | 0.458 |
| EW ICT | 0.458 | 0.458 |
| Other | 0.3 | 0.3 |

Fig. 2 (a) and (b) show the direct and LCA carbon intensity, respectively, associated with GB electricity on the selected days in 2016. We can notice that, by far, the GB grid system on the 1st December has the highest carbon intensity. The peak direct and LCA emissions recorded on 1st December at 20:00 PM are approximately 0.39 kgCO$_2$e/kWh and 0.43 kgCO$_2$e/kWh, respectively.

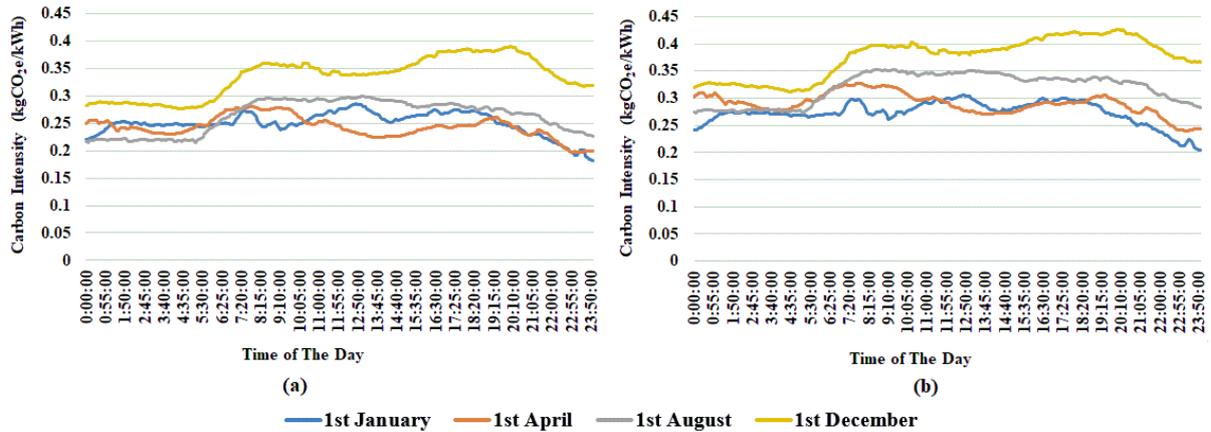

Figure 2: a) Direct and b) LCA emissions of GB electricity generation at selected days in 2016.

## 3. Energy Consumption of BT Core Network and Associated Direct and LCA emissions:

We also investigated the power consumption and 5-minutes emissions of 21st Century (21CN) BT core network, depicted in Fig. 3 which consists of 20 nodes and 68 links. Given the daily internet traffic of 2016 discussed above [3], the traffic demands in BT core network are considered to originate either from two datacenters located in London and Preston to serve distributed users (88% of the traffic), or from non-datacenter traffic between the cities (12% of the traffic). Note that the traffic demand is considered to be proportional to the population of each city [24]. Fig. 4(a) illustrates the daily network traffic in 2016 at different times of the day, which fluctuates between 6 and 14 Tbps. We investigated BT core network by formulating the problem of minimizing the total network energy consumption as a mixed integer linear programming (MILP) model. Table 2 shows the input data for the BT core network model. Power consumption estimation of BT core network in 2016 which reflects the total traffic demand is shown in Fig. 4(b).

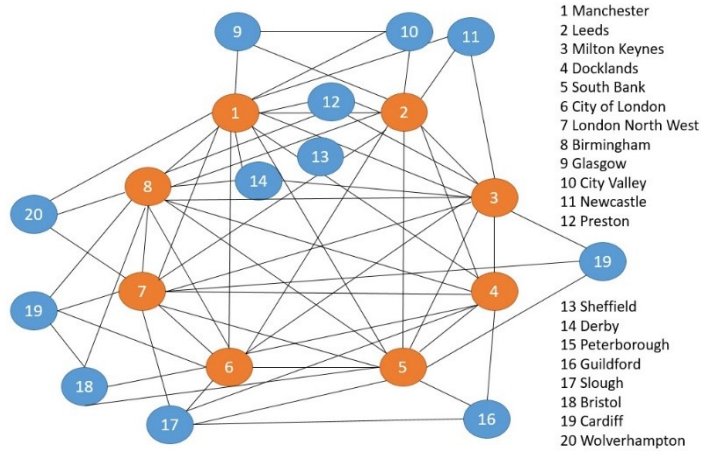

Figure 3: BT 21CN core network.

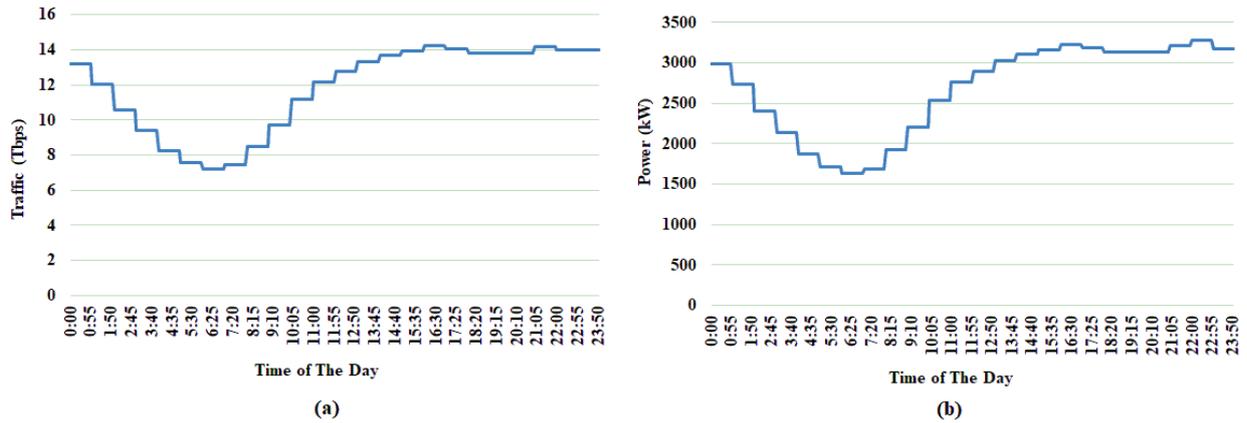

Figure 4: a) daily traffic of BT core network in 2016, b) Daily power consumption of BT core network in 2016.

*Table 2: Model Input Parameters*

| Router port power consumption | 825 W [25] |
|---|---|
| Transponder power consumption | 167 W [25] |
| Regenerator power consumption | 334 W, reach 2500 km [25] |
| EDFA power consumption | 55 W [25] |
| Optical switch power consumption | 85 W [25] |

Fig. 5(a) and (b) show the direct and LCA emissions associated with electricity production of BT core network at 5 minutes resolution, respectively. The highest daily direct and LCA emissions were recorded on 1st December are 21.8 t$CO_2$e and 24.5 t$CO_2$e, respectively.

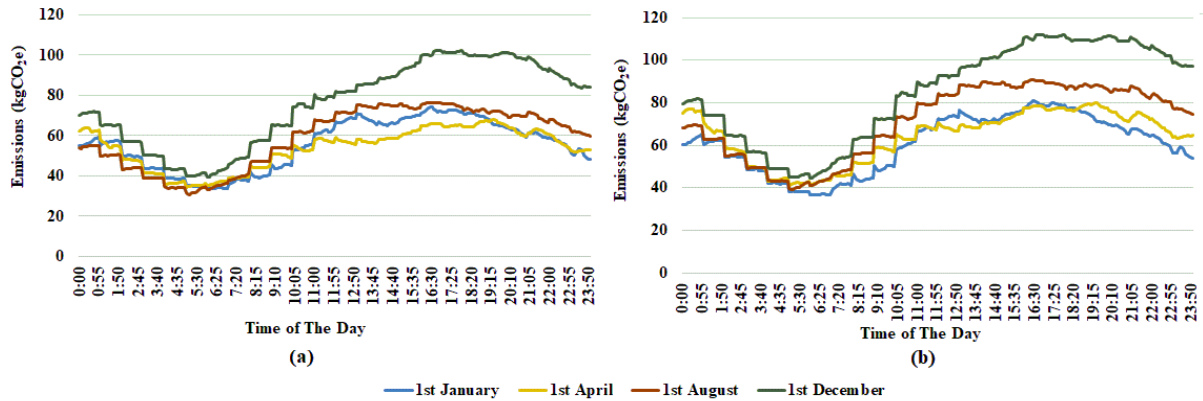

Figure 5: a) direct and b) LCA emissions associated with electricity usage of BT core network at 5 minutes resolution.

## 4. CONCLUSIONS

Real-time calculation of direct and LCA emissions of core networks considering the British Telecom (BT) network as a case study was presented in this paper. We investigated BT core network direct and LCA emissions at 5-minute resolution considering Internet traffic of 2016. Our results showed that on 1st December, BT core network emitted approximately 22 t$CO_2$e of direct emission and 25 t$CO_2$e of LCA emissions.

## ACKNOWLEDGEMENTS

The authors would like to acknowledge funding from the Engineering and Physical Sciences Research Council (EPSRC), INTERNET (EP/H040536/1) and STAR (EP/K016873/1) projects. The first author would like to acknowledge the Government of Saudi Arabia and Taibah University for funding his PhD scholarship. All data are provided in full in the results section of this paper.

## REFERENCES

[1] M. Pickavet *et al.*, "Worldwide energy needs for ICT: The rise of power-aware networking," *2008 2nd Int. Symp. Adv. Networks Telecommun. Syst. ANTS 2008*, no. December, pp. 15–17, 2008.

[2] W. Van Heddeghem, S. Lambert, B. Lannoo, D. Colle, M. Pickavet, and P. Demeester, "Trends in worldwide ICT electricity consumption from 2007 to 2012," *Comput. Commun.*, vol. 50, no. 0, pp. 64–76, 2014.

[3] Cisco System, "United Kingdom - 2021 Cisco Visual Networking Index Forecast Highlights," 2016.

[4] J. M. H. Elmirghani *et al.*, "GreenTouch GreenMeter Core Network Energy-Efficiency Improvement

[5] H. A. Alharbi, T. E. H. El-gorashi, A. Q. Lawey, and J. M. H. Elmirghani, "Energy Efficient Virtual Machines Placement in IP over WDM Networks," in *19th International Conference on Transparent Optical Networks*, 2017, pp. 1–4.

[6] M. Musa, T. Elgorashi, and J. Elmirghani, "Energy Efficient Survivable IP over WDM Networks with Network Coding," *J. Opt. Commun. Netw.*, vol. 10, no. 5, pp. 1–12, 2018.

[7] A. Q. Lawey, T. E. H. El-Gorashi, and J. M. H. Elmirghani, "Distributed energy efficient clouds over core networks," *J. Light. Technol.*, vol. 32, no. 7, pp. 1261–1281, 2014.

[8] X. Dong, T. E. H. El-Gorashi, and J. M. H. Elmirghani, "On the energy efficiency of physical topology design for IP over WDM networks," *J. Light. Technol.*, vol. 30, no. 12, pp. 1931–1942, 2012.

[9] X. Dong, T. El-Gorashi, and J. M. H. Elmirghani, "IP over WDM networks employing renewable energy sources," *J. Light. Technol.*, vol. 29, no. 1, pp. 3–14, 2011.

[10] L. Nonde, T. E. H. El-Gorashi, and J. M. H. Elmirghani, "Energy Efficient Virtual Network Embedding for Cloud Networks," *J. Light. Technol.*, vol. 33, no. 9, pp. 1828–1849, 2015.

[11] M. Musa, T. Elgorashi, and J. Elmirghani, "Energy efficient survivable IP-Over-WDM networks with network coding," *J. Opt. Commun. Netw.*, vol. 9, no. 3, pp. 207–217, 2017.

[12] X. Dong, T. El-Gorashi, and J. M. H. Elmirghani, "Green IP over WDM networks with data centers," *J. Light. Technol.*, vol. 29, no. 12, pp. 1861–1880, 2011.

[13] A. Q. Lawey, T. E. H. El-Gorashi, and J. M. H. Elmirghani, "BitTorrent Content Distribution in Optical Networks," *J. Light. Technol.*, vol. 32, no. 21, pp. 4209–4225, 2014.

[14] G. S. G. Shen and R. S. Tucker, "Energy-Minimized Design for IP Over WDM Networks," *IEEE/OSA J. Opt. Commun. Netw.*, vol. 1, no. 1, pp. 176–186, 2009.

[15] N. I. Osman, T. El-Gorashi, L. Krug, and J. M. H. Elmirghani, "Energy-efficient future high-definition TV," *J. Light. Technol.*, vol. 32, no. 13, pp. 2364–2381, 2014.

[16] INTERNATIONAL ENERGY AGENCY, "Trends $CO_2$ Emissions," 2015. [Online]. Available: https://www.iea.org/publications/freepublications/publication/CO2EmissionsTrends.pdf. [Accessed: 30-Mar-2018].

[17] R. Turconi, A. Boldrin, and T. Astrup, "Life cycle assessment (LCA) of electricity generation technologies: Overview, comparability and limitations," *Renew. Sustain. Energy Rev.*, vol. 28, pp. 555–565, 2013.

[18] National Grid, "Carbon Intensity Forecast Methodology," 2017.

[19] S. Soimakallio, J. Kiviluoma, and L. Saikku, "The complexity and challenges of determining GHG (greenhouse gas) emissions from grid electricity consumption and conservation in LCA (life cycle assessment) - A methodological review," *Energy*, vol. 36, no. 12, pp. 6705–6713, 2011.

[20] "Elexon Portal." [Online]. Available: https://www.bmreports.com/bmrs/?q=eds/main. [Accessed: 01-Apr-2018].

[21] "G.B. National Grid Status." [Online]. Available: http://www.gridwatch.templar.co.uk/. [Accessed: 01-Apr-2018].

[22] B. W. Ang and B. Su, "Carbon emission intensity in electricity production: A global analysis," *Energy Policy*, vol. 94, pp. 56–63, 2016.

[23] S. Schlomer *et al.*, "Annex III: Technology-Specific Cost and Performance Parameters," *Clim. Chang. 2014 Mitig. Clim. Chang. Contrib. Work. Gr. III to Fifth Assess. Rep. Intergov. Panel Clim. Chang.*, pp. 1329–1356, 2014.

[24] Office for National Statistics, "Population estimates for UK, England and Wales, Scotland and Northern Ireland: mid-2016." [Online]. Available: https://www.ons.gov.uk/peoplepopulationandcommunity/populationandmigration/populationestimates/bulletins/annualmidyearpopulationestimates/mid2016. [Accessed: 01-Apr-2018].

[25] "GreenTouch Green Meter Research Study : Reducing the Net Energy Consumption in Communications Networks by up to 90 % by 2020," *White Pap.*, pp. 1–25, 2015.